\documentclass[12pt,english]{article}
\pdfoutput=1
\usepackage{graphicx,array}
\usepackage{cite}
\usepackage{color}
\usepackage{latexsym}
\usepackage{amsthm}
\usepackage{amsmath}
\usepackage[titletoc]{appendix}
\usepackage{enumitem}
\usepackage{amssymb}
\usepackage{color}
\usepackage[unicode=true,
 bookmarks=true,bookmarksnumbered=false,bookmarksopen=false,
 breaklinks=false,pdfborder={0 0 1},backref=false,colorlinks=true]
 {hyperref}
\hypersetup{pdftitle={Higher-Point Positivity},
 pdfauthor={Venkatesa Chandrasekaran, Arvin Shahbazi Moghaddam, Grant N. Remmen},
 citecolor=black,linkcolor=black,urlcolor=black}
\usepackage{breakurl}
\usepackage[hang,flushmargin]{footmisc} 

\def\simgt{\mathrel{\lower2.5pt\vbox{\lineskip=0pt\baselineskip=0pt
           \hbox{$>$}\hbox{$\sim$}}}}
\def\simlt{\mathrel{\lower2.5pt\vbox{\lineskip=0pt\baselineskip=0pt
           \hbox{$<$}\hbox{$\sim$}}}}

\newcommand{\be}{\begin{equation}}
\newcommand{\ee}{\end{equation}}
\newcommand{\bea}{\begin{eqnarray}}
\newcommand{\eea}{\end{eqnarray}}
\newcommand{\Eq}[1]{Eq.~(\ref{#1})}
\newcommand{\Eqs}[2]{Eqs.~(\ref{#1}) and (\ref{#2})}
\newcommand{\Sec}[1]{Sec.~\ref{#1}}
\newcommand{\Secs}[2]{Secs.~\ref{#1} and \ref{#2}}

\newcommand{\Ref}[1]{Ref.~\cite{#1}}

\newcommand{\ket}[1]{\left| #1 \right\rangle}

\newcommand*\oline[1]{%
  \vbox{%
    \hrule height 0.5pt
    \kern0.68ex
    \hbox{%
      \kern-0.1em
      \ifmmode#1\else\ensuremath{#1}\fi
      \kern-0.1em
    }
  }
}

\definecolor{nicered}{rgb}{0.7,0.1,0.1}
\definecolor{nicegreen}{rgb}{0.1,0.5,0.1}
\hypersetup{
 colorlinks,citecolor=black,linkcolor=black,urlcolor=black}
\usepackage{xcolor}

\begin{document}

\interfootnotelinepenalty=10000
\baselineskip=18pt

\hfill\hfill

\vspace{2cm}
\thispagestyle{empty}
\begin{center}
{\LARGE \bf
Higher-Point Positivity
}\\
\bigskip\vspace{1cm}{
{\large Venkatesa Chandrasekaran, Grant N. Remmen, \\and Arvin Shahbazi-Moghaddam}
} \\[7mm]
 {\it Center for Theoretical Physics and Department of Physics\\
    University of California, Berkeley, CA 94720, USA and \\
Lawrence Berkeley National Laboratory, Berkeley, CA 94720, USA} 
\let\thefootnote\relax\footnote{e-mail:\\
\url{ven_chandrasekaran@berkeley.edu}, \url{grant.remmen@berkeley.edu}, \url{arvinshm@berkeley.edu}}
 \end{center}
\bigskip
\centerline{\large\bf Abstract}

\begin{quote} \small
We consider the extension of techniques for bounding higher-dimension operators in quantum effective field theories to higher-point operators. Working in the context of theories polynomial in $X=(\partial \phi)^2$, we examine how the techniques of bounding such operators based on causality, analyticity of scattering amplitudes, and unitarity of the spectral representation are all modified for operators beyond $(\partial \phi)^4$. Under weak-coupling assumptions that we clarify, we show using all three methods that in theories in which the coefficient $\lambda_n$ of the $X^n$ term for some $n$ is larger than the other terms in units of the cutoff, $\lambda_n$ must be positive (respectively, negative) for $n$ even (odd), in mostly-plus metric signature. Along the way, we present a first-principles derivation of the propagator numerator for all massive higher-spin bosons in arbitrary dimension. We remark on subtleties and challenges of bounding $P(X)$ theories in greater generality. Finally, we examine the connections among energy conditions, causality, stability, and the involution condition on the Legendre transform relating the Lagrangian and Hamiltonian.
\end{quote}
	
\setcounter{footnote}{0}

\newpage
\tableofcontents
\newpage

\section{Introduction}\label{sec:Introduction}

A dramatic development in our knowledge of quantum field theory has been the discovery that not all effective field theories are consistent with ultraviolet completion in quantum gravity. Certain Lagrangians that one can write down possess pathologies that are a priori hidden, but that can be elucidated though careful consideration of consistency conditions that can be formulated in the infrared and that are thought to be obeyed by any reasonable ultraviolet completion. Such infrared conditions include analyticity of scattering amplitudes, quantum mechanical unitarity, and causality of particle propagation \cite{Adams:2006sv,Jenkins:2006ia,Gruzinov:2006ie,Dvali:2012zc,Cheung:2014ega,Bellazzini:2014waa,Bellazzini:2015cra,Cheung:2016yqr,Cheung:2016wjt,Bellazzini:2016xrt,Nicolis:2009qm,Aharonov:1969vu,Camanho:2014apa}, as well as self-consistency of black hole entropy in the context of the recent proof of the weak gravity conjecture \cite{Cheung:2018cwt}. Delineating the space of consistent low-energy effective field theories is of great current interest in the context of the swampland program \cite{Vafa:2005ui,Ooguri:2006in,ArkaniHamed:2006dz}, which seeks to characterize and bound in theory space the possible effective field theories amenable to ultraviolet completion in quantum gravity. Infrared requirements form a powerful set of tools, giving us rigorous positivity bounds that complement intuition from ultraviolet examples. Such self-consistency constraints have been used to bound the couplings of many different higher-dimension operators in scalar field theory \cite{Adams:2006sv}, gauge theory \cite{Adams:2006sv}, Einstein-Maxwell theory \cite{Cheung:2014ega,Cheung:2018cwt}, higher-curvature corrections to gravity \cite{Gruzinov:2006ie,Bellazzini:2015cra,Cheung:2016wjt}, and massive gravity \cite{Cheung:2016yqr}.

The simplest positivity bound on effective theories applies to the coupling of the $(\partial \phi)^4$ operator. In a massless theory of a real scalar $\phi$ with a shift symmetry, the first higher-dimension operator that one can add to the kinetic term $-\partial_\mu \phi \partial^\mu \phi/2$ is the operator
\be
(\partial \phi)^4 = \partial_\mu \phi \partial^\mu \phi \partial_\nu \phi \partial^\nu \phi. 
\ee
In a theory given by $-\frac{1}{2}(\partial \phi)^2 + \lambda (\partial \phi)^4$, the forward amplitude for two-to-two $\phi$ scattering is ${\cal A}(s) = 16\lambda s^2$. A standard dispersion relation argument \cite{Adams:2006sv} then relates the coefficient of $s^2$ in this forward amplitude at low energies to an integral over the cross section at high energies, which physically must be positive. That is, analyticity of scattering amplitudes guarantees that $\lambda$ is positive. Similarly, one can compute the speed of propagation of $\phi$ perturbations in a nonzero $\phi$ background: one finds that subluminality requires $\lambda > 0$ and that if $\lambda < 0$ it is straightforward to build causal paradoxes involving superluminal signaling between two bubbles of $\phi$ background with a relative boost. A litany of other examples of analyticity and causality bounds focuses on similar four-point interactions, though for more complicated theories and fields involving gauge bosons and gravitons. 

In this paper, we explore a new direction in the space of positivity bounds: higher-point operators. In particular, we will bound the $P(X)$ theory, whose Lagrangian is simply a polynomial in 
\be 
X = \partial_\mu \phi \partial^\mu \phi,
\ee
which in the effective field theory we can write as\footnote{We will use mostly-plus metric signature throughout.} 
\be 
{\cal L} = -\frac{1}{2} X + \sum_{i=2}^\infty \lambda_i X^i.\label{eq:L}
\ee 

A case of particular tractability is an $n$th-order $P(X)$ theory, in which the $\lambda_i$ are very small or zero for $i < n$ for some $n>1$, where $n$ is the first nonnegligible higher-order term in the $P(X)$ polynomial:
\be 
{\cal L} = -\frac{1}{2}X + \sum_{i = n}^\infty \lambda_i X^i.\label{eq:delayed}
\ee 
We use a weak-coupling assumption from the ultraviolet to the infrared to guarantee a well-defined $\hbar$ counting at all energy scales, as in \Ref{Cheung:2016wjt}, so that the vanishing of the tree-level $\lambda_i$ for $i<n$ is well defined.

We will show that analyticity of scattering amplitudes and causality of signal propagation imply the same positivity bound on the theory in \Eq{eq:delayed}:
\be
\begin{aligned}
\lambda_n > 0 & \text{ if $n$ is even},\\
\lambda_n < 0 & \text{ if $n$ is odd}.
\end{aligned} \label{eq:bd}
\ee
We will also find that \Eq{eq:bd} comes about as a consequence of unitarity of quantum mechanics in the context of spectral representations for a particular class of ultraviolet completions. This bound represents progress for the program of constraining the allowed space of self-consistent low-energy effective theories, constituting a generalization of the well known $(\partial \phi)^4$ bound. Further, the formalism we develop along the way for applying infrared consistency bounds to higher-point operators is useful in its own right.

Considering $X^n$ as the first nonnegligible operator in the effective field theory can be motivated physically in several different ways. We can consider tree-level completions of the $X^i$ operators through massive states coupling to $(\partial \phi)^i$.  If there is no coupling of massive states to $(\partial \phi)^i$ for $i<n$, then the tree-level value of $\lambda_i$ vanishes for $i<n$. We can then place the positivity bound in \Eq{eq:bd} on $\lambda_n$ using the tree-level amplitude. Note that this logic does not contradict the positivity bound on $(\partial \phi)^4$ in \Ref{Adams:2006sv}, since $\lambda_2$ could still be generated at loop level, though $\lambda_n$ from the tree-level completion would be parametrically larger in units of the cutoff.\footnote{We assume a sufficiently weak coupling that it is consistent to drop the lower-point operators that are suppressed by loop factors, despite additional bounds coming from inelastic scattering \cite{Bellazzini:2017fep}.} Moreover, from the perspective of the effective field theory, the higher-dimension operators in the $n$th-order $P(X)$ theory in \Eq{eq:delayed} can be viewed as a sector of a larger theory. For example, taking a complex scalar $\phi$ with a $\mathbb{Z}_n$ symmetry $\phi\rightarrow e^{2\pi i m/n} \phi$ for integer $m$, the allowed higher-dimension operators are of the form $X^{np}$, $\bar{X}^{np}$, and $\hat{X}^p$ for integer $p$, where $\bar{X} = \partial_\mu \phi^* \partial^\mu \phi^*$ and $\hat X = \partial_\mu \phi \partial^\mu \phi^*$. In particular, all operators $X^i$ for $i<n$ would be forbidden and the scattering of $2n$ $\phi$ particles at tree level would occur only through the $X^n$ contact operator, just as in the $n$th-order $P(X)$ theory in \Eq{eq:delayed}.

This paper is organized as follows. In \Sec{sec:Analyticity}, we consider the application of analyticity bounds for higher-point amplitudes and derive our bound \eqref{eq:bd} on the $n$th-order $P(X)$ theory. Next, in \Sec{sec:Causality} we find that  the bound \eqref{eq:bd} also follows from demanding the absence of causal paradoxes. In \Sec{sec:Unitarity} we consider a particular class of tree-level completions and find that the couplings obey \Eq{eq:bd} as a consequence of unitarity of the spectral representation. Along the way, we present an elegant derivation of the propagator for higher-spin massive bosons in arbitrary spacetime dimension. We discuss the obstacles, in the form of kinematic singularities, that preclude  straightforward generalization of some of these bounds to arbitrary (i.e., not strictly $n$th-order) $P(X)$ theories in \Sec{sec:General}. In \Sec{sec:Legendre} we show that there is a deep relationship between positivity bounds and the involution property of the Legendre transform relating the Lagrangian and Hamiltonian formulations of the mechanics of the $P(X)$ theory. We conclude and discuss future directions in \Sec{sec:Conclusions}.

\section{Bounds from Analyticity}\label{sec:Analyticity}

In this section, we derive the bound in \Eq{eq:bd} through a generalization of the dispersion relation argument that has been previously applied to two-to-two scattering amplitudes \cite{Adams:2006sv}. We first discuss formalism for general $n$-to-$n$ particle scattering, before considering our specific theory of interest and deriving the bounds.

\subsection{The Forward Limit}

Consider a general effective field theory for which one wishes to bound the couplings of higher-dimension operators using analyticity of scattering amplitudes. Fundamentally, such positivity bounds come from the optical theorem, ${\rm Im}\,{\cal A}(s)= s \sigma(s)$ where ${\cal A}(s)$ is the forward amplitude, $s$ is the center-of-mass energy of the incoming particles, and $\sigma$ is the cross section, which is mandated physically to be positive. Taking a four-point operator, kinematics allows only one forward limit (module polarization or other, internal degrees of freedom):
\be
\begin{aligned}
p_3 &= -p_1 \\
p_4 &= -p_2,
\end{aligned} 
\ee
working in the convention of all momenta incoming. 

However, at higher-point, there are multiple forward kinematic configurations, given by the angles that the various momenta make with respect to each other. In particular, considering $n$-to-$n$ particle scattering, going to forward kinematics so that $p_{n+i} = -p_i$ for $1\leq i \leq n$, there is a family of forward limits parameterized by $(D-2)\times (n-2)$ independent angles and $n-1$ independent energies.
The reason for this counting is as follows. A priori, we choose an angle on the celestial sphere for the direction associated for each of the $p_i$, $1\leq i \leq n$. Momentum is conserved automatically by the forward condition. Moreover, we can use Lorentz invariance to fix two of the directions: one angle is fixed by rotational invariance and another is fixed by boost symmetry, which allows us to take two of the pairs to be back-to-back with equal energy. Hence, we can fix $n-2$ points on the celestial sphere, each of which requires $D-2$ angular coordinates in $D$ spacetime dimensions.

This large number of possible forward limits means that higher-point amplitudes have significant power to constrain the couplings of higher-point operators, despite the larger number of operators one can write down.

\subsection{Higher-Point Dispersion Relations and Bounds for $P(X)$}\label{sec:dispersion}

Placing positivity bounds using higher-point amplitudes follows a generalization of the argument bounding four-point operators. First, let us define the Mandelstam invariants 
\be 
s_{ij} = -(p_i + p_j)^2 = -2p_i \cdot p_j. 
\ee
There are $n(2n-3)$ independent Mandelstam invariants for the $2n$-point amplitude (i.e., $n$-to-$n$ scattering), taking into account momentum conservation and the on-shell conditions. Choosing a particular forward limit, by fixing all $(D-2)\times(n-2)$ of the angular parameters, $\cal A$ becomes a function of the remaining nonzero $s_{ij}$. In particular, we will choose as our variable for analytic continuation the center-of-mass energy squared,
\be 
s = -(p_1 + \cdots + p_n)^2.
\ee

We wish to place a bound on the couplings of the $n$th-order $P(X)$ theory \eqref{eq:delayed} for even or odd $n$, where the first nonnegligible $\lambda_i$ coefficient of the $X^i$ operator occurs at $i=n$. Making particular concrete choices for the kinematics will allow us to bound the coefficient $\lambda_n$. We will find that different choices of kinematics and dispersion relations are needed for $n$ even or odd.

At general kinematics, the $2n$-point tree-level amplitude for the $n$th-order $P(X)$ theory is 
\be
{\cal A} = \frac{\lambda_n}{2^n} \sum_{\{\sigma\}} s_{\sigma_1 \sigma_2} \cdots s_{\sigma_{2n-1}\sigma_{2n}},\label{eq:A_contact}
\ee
where $\sigma$ runs over the the $(2n)!2^{-n}$ different possible groupings of $\{1,\ldots,2n\}$ into an ordered list of $n$ unordered pairs. Throughout this section, we will work with a weak-coupling assumption from the infrared to the deep ultraviolet, above the cutoff, implying a well defined $\hbar$ expansion at all scales \cite{Cheung:2016wjt}. For our $n$th-order $P(X)$ theory, such an assumption will allow us to ignore disconnected components of the amplitude in the generalized optical theorem, since the loop contributions to the disconnected amplitude will be negligible and the tree-level components will vanish except for the contact diagram.

\subsubsection{Even $n$}\label{sec:even}

If $n$ is even, we choose the following forward kinematics:
\be
\begin{aligned}
p_i &= p_1 &\text{\qquad for \qquad}& i=1\mod 2\\ p_i &= p_2 &\text{\qquad for \qquad}& i=0\mod 2\\ p_{i+n} &= -p_i\label{eq:pchoiceeven}
\end{aligned}
\ee
for all $i$, $1\leq i\leq n$. Then the center-of-mass energy is
\be
s = \frac{n^2}{4} s_{12}\label{eq:seven}
\ee
and the forward amplitude, within the regime of our weakly-coupled effective field theory, is
\be
{\cal A}(s) = (n!)^2 \left(\frac{2}{n}\right)^{2n} \lambda_n  s^{n}.\label{eq:A_contact_even}
\ee

In the complex $s$ plane, we consider the contour integral
\be
I_n = \frac{1}{2\pi i} \oint_\gamma \frac{\mathrm{d}s}{s^{n+1}} {\cal A} (s) = (n!)^2 \left(\frac{2}{n}\right)^{2n} \lambda_n,\label{eq:In}
\ee
where $\gamma$ is a small contour around the origin. Similarly, we can define
\be 
I'_n = \frac{1}{2\pi i} \oint_{\gamma'} \frac{\mathrm{d}s}{s^{n+1}} {\cal A} (s),\label{eq:I'n}
\ee
where $\gamma'$ is a contour running just above and below the real $s$ axis, plus a boundary contour at infinity.

The standard analyticity assumptions of the S-matrix imply that ${\cal A}$ is analytic everywhere except for poles in the $s_{ij}$ where massive states in the ultraviolet completion go on-shell and, at loop level, branch cuts associated with massive states in loops. See Refs.~\cite{Elvang:2012st,Logunov:1977xb} for a discussion of analyticity for 3-to-3 scattering and \Ref{Eden:1966dnq} for a more general treatment of the analytic S-matrix. (If we made the more restrictive assumption of a tree-level ultraviolet completion, then the nonanalyticities would only occur at the poles of the massive states, as one could see by explicit construction of the Feynman diagrams.) Given the choice of kinematics in \Eq{eq:pchoiceeven}, the only independent nonzero $s_{ij}$ is $s_{12}$, which is equivalent to a rescaled version of $s$ by \Eq{eq:seven}. Hence, all nonanalyticities in the complex $s$ plane occur at a set of poles  (and branch cuts) on the real $s$ axis. That is, Cauchy's theorem implies that $I_n = I'_n$. We assume that the boundary integral at infinity vanishes. For a massive theory,  this would follow from the Froissart bound $|{\cal A}(s) < |s \log^{D-2} s|$ at large $|s|$ \cite{Chaichian:1987zt,Chaichian:1992hq}. Even though we are considering a massless theory, it is reasonable to assume some form of polynomial boundedness that forbids the amplitude from diverging too quickly with $s$ at large $s$; in essence, discarding the boundary integral is equivalent to demanding that the $X^n$ term in the action is in fact ultraviolet completed, i.e., forbidding primordial $X^n$ terms by demanding that the higher-dimension operator originate from the exchange of states at some scale.

Equating $I_n = I'_n$, we thus have
\be
\lambda_n = \frac{1}{2\pi i(n!)^2}\left(\frac{n}{2}\right)^{2n} \left(\int_{-\infty}^{-s_0} + \int_{s_0}^\infty\right) \frac{\mathrm{d}s}{s^{n+1}} \,\mathrm{disc}\,{\cal A}(s),
\ee
where $s_0$ is some regulator below which we take the amplitude to be analytic and $\mathrm{disc}\,{\cal A}(s) = {\cal A}(s+i\epsilon) - {\cal A}(s-i\epsilon)$. For example, if we use $\hbar$ counting to restrict to the tree-level scattering amplitude, we can take $s_0$ to be of order the scale of the ultraviolet completion.

In the two-to-two scattering case, the integrals over the positive and negative real $s$ axis are related by the  crossing symmetry associated with swapping $p_1$ and $p_3$, i.e., by swapping the $s$ and $u=-s-t$ channels for forward kinematics. For our present calculation involving $n$-to-$n$ scattering, crossing symmetry implies that the amplitude is invariant under swapping legs $n$ and $2n$. With the choice of kinematics in \Eq{eq:pchoiceeven}, this is equivalent to swapping legs $p_i$ for $p_{i+n}$ for all even $i$ between $2$ and $n$, which has the effect of swapping $p_2 \leftrightarrow -p_2$ while leaving $p_1$ unchanged, so $s_{12}\leftrightarrow -s_{12}$ and $s\leftrightarrow -s$. Hence, as in the two-to-two case, crossing symmetry implies that with our choice of kinematics ${\cal A}(s)$ is an even function of $s$, even in the ultraviolet. We thus have ${\rm disc}\,{\cal A}(-s) = -{\rm disc}\,{\cal A}(s)$ and
\be
\lambda_n = -\frac{i}{\pi (n!)^2}\left(\frac{n}{2}\right)^{2n} \int_{s_0}^\infty\frac{\mathrm{d}s}{s^{n+1}}\, \mathrm{disc}\,{\cal A}(s).
\ee

Using the Schwarz reflection principle ${\cal A}(s^*) = [{\cal A}(s)]^*$, we have ${\rm disc}\,{\cal A}(s)=2i\,{\rm Im}\,{\cal A}(s)$. In two-to-two scattering, the optical theorem relates the cross-section to the imaginary part of the forward amplitude. Generalized to an initial multiparticle state $\ket{n,s}$ with center-of-mass energy $s$, the optical theorem implies  
\be
{\rm Im}\,{\cal A}(s) = \frac{1}{2}\sum_{X} \int {\rm dLIPS}_{X}|{\cal A}(\ket{n,s}\rightarrow X)|^2,\label{eq:dLIPS}
\ee
where the sum is over all intermediate states $X$, ${\rm dLIPS}_X=(2\pi)^D \delta^D(\sum_{i=1}^n p_i + p_X)\Pi_{j\in X}\frac{{\rm d}^{D-1} \vec{p}_j}{(2\pi)^{D-1}}\frac{1}{2E_j}$ is the Lorentz-invariant phase space measure for the intermediate state \cite{Schwartz:2013pla}, and ${\cal A}(\ket{n,s}\rightarrow X)$ is the amplitude for the $n$-particle initial state with center-of-mass energy $s$ going to the final state $X$. Note that, in general, the appropriate amplitude appearing in the generalized optical theorem \eqref{eq:dLIPS} is the full $n$-to-$n$ amplitude, including disconnected diagrams \cite{Weinberg:1995mt}. However, as noted above, for the $n$th-order $P(X)$ theory we consider, the only contribution to the $n$-point amplitude comes from the tree-level diagrams. Hence, for the theory at hand, \Eq{eq:dLIPS} applies to the connected component of the amplitude alone.

The right-hand side of \Eq{eq:dLIPS} is manifestly positive. Thus, we have a bound on $\lambda_n$ in the $n$th-order $P(X)$ theory for even $n$:
\be
\lambda_n = \frac{1}{\pi (n!)^2}\left(\frac{n}{2}\right)^{2n} \int_{s_0}^\infty \frac{\mathrm{d}s}{s^{n+1}} \sum_{X} \int {\rm dLIPS}_{X}|{\cal A}(\ket{n,s}\rightarrow X)|^2> 0. 
\ee

\subsubsection{Odd $n$}
For the $n$th-order $P(X)$ theory where $n$ is odd, we choose the kinematics
\be
\begin{aligned}
p_i &= p_1 &\text{\qquad for \qquad}& i=1\mod 2,\; i\in[1,n-1] \\ p_i &= p_2 &\text{\qquad for \qquad}& i=0\mod 2,\; i\in[1,n-1] \\ p_{i+n} &= -p_i &\text{\qquad for \qquad}& i\in [1,n].
\end{aligned}
\ee
With these choices of kinematics, we have the center-of-mass energy
\be
s = \frac{(n-1)^2}{4}s_{12} + \frac{n-1}{2}s_{1n} + \frac{n-1}{2} s_{2n}.
\ee
and the forward amplitude, in our weakly-coupled low-energy effective field theory, is
\be
{\cal A} =-2 (n-1)n!(n-1)! \lambda_n s_{12}^{n-2} s_{1n}s_{2n}.
\ee

We can make a further choice of kinematics to set $s_{1n} = s_{2n}$, which we will for brevity call $s_n$, and analytically continue in $s_{n}$ while holding $\delta = (n-1)^2 s_{12}/4$ constant. That is, the center-of-mass energy is $s = (n-1)s_n + \delta$, so analytic continuation in $s$ is equivalent to analytic continuation in $s_n$.\footnote{In \Ref{Elvang:2012st}, a related choice of kinematics was made for six-point scattering, in a dilaton effective action relevant for the $a$-theorem in $D=6$.} Note that for physical kinematics, $\delta > 0$. The forward amplitude is
\be 
{\cal A}(s) =-2^{2n-3} (n-1)^{-2n+5} n!(n-1)!  \delta^{n-2}  \lambda_n  s_n^2 = -\left(\frac{2}{n-1}\right)^{2n-3} n!(n-1)!\delta^{n-2}  \lambda_n (s-\delta)^2.
\ee 

In contrast with \Sec{sec:even}, we define the contour integrals for odd $n$ as
\be
I_n = \frac{1}{2\pi i} \oint_\gamma \frac{\mathrm{d}s}{s^{3}} {\cal A} (s) = -\left(\frac{2}{n-1}\right)^{2n-3} n!(n-1)!\delta^{n-2}  \lambda_n ,\label{eq:Ino}
\ee
for a small contour $\gamma$ around the origin and
\be 
I'_n = \frac{1}{2\pi i} \oint_{\gamma'} \frac{\mathrm{d}s}{s^{3}} {\cal A} (s)=\frac{1}{2\pi i}\left(\int_{-\infty}^{-s_0} + \int_{s_0}^\infty\right) \frac{\mathrm{d}s}{s^{3}}\, \mathrm{disc}\,{\cal A}(s),\label{eq:I'no}
\ee
for a contour $\gamma'$ running just above and below the real $s$ axis, plus a boundary contour at infinity that we drop as before.

Crossing symmetry under swapping legs $n$ and $2n$ is equivalent under our choice of kinematics to swapping $p_{n} \leftrightarrow p_{2n} = -p_{n}$, i.e., swapping $s_n \leftrightarrow -s_n$ while holding $s_{12}$ (and thus $\delta$) fixed. That is, the forward amplitude, even in the ultraviolet, must be an even function of $s_n$. Equivalently, the full forward amplitude satisfies
\be
\mathcal{A}(s) = \mathcal{A}(-s+2 \delta).
\ee 
We therefore have
\be
\begin{aligned}
{\rm disc}\, \mathcal{A}(-s) &= \mathcal{A}(-s+i \epsilon) - \mathcal{A}(-s - i\epsilon)\nonumber\\
&=\mathcal{A}(s-i \epsilon+2 \delta) - \mathcal{A}(s+i \epsilon+2 \delta) \\&= -{\rm disc}\, \mathcal{A}(s+2 \delta).
\end{aligned}
\ee
Using analyticity to equate $I_n$ and $I'_n$ in \Eqs{eq:Ino}{eq:I'no} and using the Schwarz reflection principle and the optical theorem as before, we obtain a bound on $\lambda_n$ in the $n$th-order $P(X)$ theory for odd $n$:
\be 
\begin{aligned}
\lambda_n &= -\frac{1}{2\pi i\, n!(n-1)!\delta^{n-2}}\left(\frac{n-1}{2}\right)^{2n-3} \int_{s_0}^{\infty}\frac{{\rm d}s}{s^3} [{\rm disc}\,{\cal A}(s) + {\rm disc}\,{\cal A}(s+2\delta)] \\
&=-\frac{1}{\pi\, n!(n-1)!\delta^{n-2}}\left(\frac{n-1}{2}\right)^{2n-3}  \int_{s_0}^{\infty} \frac{{\rm d}s}{s^3}[{\rm Im}\,{\cal A}(s) + {\rm Im}\,{\cal A}(s+2\delta)] \\
&=-\frac{1}{2\pi \, n!(n-1)!\delta^{n-2}}\left(\frac{n-1}{2}\right)^{2n-3} \int_{s_0}^{\infty}\frac{{\rm d}s}{s^3}  \left[\sum_{X} \int {\rm dLIPS}_{X}|{\cal A}(\ket{n,s}\rightarrow X)|^2\right. \\ & \left. \hspace{7.7cm} + \sum_{X} \int {\rm dLIPS}_{X}|{\cal A}(\ket{n,s+2\delta}\rightarrow X)|^2\right] \\& < 0.
\end{aligned}
\ee

\section{Bounds from Causality}\label{sec:Causality}

Next, let us consider how bounds on the $P(X)$ theory can be derived from causality. For now, we will consider an arbitrary $P(X)$ theory, with no assumptions about the relative sizes of the various higher-dimension operators. The equation of motion for this theory is:
\be
\partial_\mu (P'(X) \partial^\mu \phi) = 0,\label{eq:eombkg}
\ee
which is solved by a constant background $\phi$ condensate, $\partial_\mu \phi = w_\mu = \text{constant}$. We will use bars to denote background vaues of fields, so $\overline{\partial_\mu \phi} = w_\mu$ and $\overline{X} = w^2$. The leading-order action for the fluctuation $\varphi = \phi - \bar \phi$ can be written as 
\be 
\mathcal{L}_\varphi = -\frac{1}{2}\tilde{\eta}^{\mu\nu}\partial_\mu \varphi \partial_\nu \varphi,
\ee
where 
\be 
\tilde{\eta}_{\mu\nu} = - \left.\frac{\delta^2 \mathcal{L}}{\delta(\partial^\mu \phi) \,\delta(\partial^\nu \phi)}\right|_{\phi = \bar \phi} = -2P'(w^2) \eta_{\mu\nu} - 4P''(w^2)w_\mu w_\nu.\label{eq:effectivemetric}
\ee
The term in the action zeroth-order in $\varphi$ is a cosmological constant $P(w^2)$, which can be dropped, while the term first-order in $\varphi$ is a tadpole, which vanishes because $\bar \phi$ satisfies the background equations of motion \eqref{eq:eombkg}.

Let us compute the speed of propagation for fluctuations about this background. The equation of motion for $\varphi$ in the $\bar \phi$ background is 
\be
\tilde{\eta}^{\mu\nu} \partial_\mu \partial_\nu \varphi = 0.
\ee
Taking a plane-wave ansatz for $\varphi$, we have the dispersion relation $\tilde{\eta}^{\mu\nu} k_\mu k_\nu = 0$, that is,
\be
P'(w^2)k^2 + 2 P''(w^2) (k\cdot w)^2 = 0. 
\ee
Writing $k_\mu = (k_0,\vec k)$, the speed of propagation is $v = k_0/|\vec k|$, which satisfies
\be
P'(w^2)(-v^2 + 1)+2P''(w^2)(-v w_0 + \hat{k}\cdot \vec w)^2 = 0, 
\ee
where $\hat{k} = \vec k/|\vec k|$ and $w_\mu = (w_0,\vec w)$. We note that $(-v w_0 + \hat{k}\cdot \vec w)^2$ is always nonnegative and can be chosen to be strictly positive for nonzero $w$ by choosing the direction of $\hat{k}$. Moreover, we choose $w$ so that $P'(w^2)$ is nonzero. It follows that $v \leq 1$ if and only if
\be
\frac{P''(w^2)}{P'(w^2)} \leq 0. 
\ee

If we are to impose a causality condition on the fluctuations $\varphi$, to be conservative we should for consistency impose a similar condition on the background itself. That is, we should require that the background energy-momentum not propagate faster than light. The relevant energy condition mandating this causal flow of energy-momentum is the null dominant energy condition (NDEC) \cite{Carroll:2003st}, which is the statement of the null energy condition (NEC), $T_{\mu\nu}\ell^\mu \ell^\nu \geq 0$ for all null $\ell^\mu$, along with the requirement that $T_{\mu\nu}\ell^\nu$ be timelike or null. The background energy-momentum tensor is
\be 
\overline{T}_{\mu\nu} = P(w^2)\eta_{\mu\nu}-2 P'(w^2) w_\mu w_\nu,\label{eq:energymomentum}
\ee
so the NEC implies  $P'(w^2)\leq 0$. Hence, we conclude that
\be
P''(w^2) \geq 0\label{eq:causalbound}
\ee
in order to guarantee $v \leq 1$ (see also \Ref{Adams:2006sv}). As shown in Refs.~\cite{Adams:2006sv,Cheung:2014ega}, if $v>1$ one can immediately form a causal paradox by highly boosting two bubbles of background condensate relative to each other in an otherwise empty region of space; sending superluminal signals back and forth between the two forms a closed signal trajectory in spacetime.

In addition to the NEC, the NDEC implies that $w^\mu$ is causal (i.e., timelike or null). The reason for this is as follows. Suppose that $w$ is spacelike, $w^2>0$. Then, defining $u_\mu = \overline{T}_{\mu\nu}\ell^\nu$ for some null $\ell$, we have
\be 
\begin{aligned}
u^2 =4(\ell\cdot w)^2 P'(w^2) \left[- P(w^2)+  P'(w^2)w^2\right].
\end{aligned}
\ee
Now, taking $P(X)$ to contain no cosmological constant, we have $P(0)=0$, so since $P'(w^2)\leq 0$, it follows that $P(w^2)\leq 0$ for $w^2>0$. Moreover, since we are interested in an interacting theory, $P''(w^2)\neq 0$ for $w^2>0$, so $P''(w^2)>0$, $P'(w^2)<0$, and $P(w^2)<0$. As a result, $P'(w^2)w^2 < P(w^2)$ for $w^2 > 0$ and, since we can choose the orientation of $\ell$ so that $\ell \cdot w \neq 0$, we have $u^2 > 0$. That is, $\overline{T}_{\mu\nu}\ell^\nu$ is spacelike, contradicting the NDEC. We conclude that $w^2$ cannot be positive, so $w^\mu$ is causal.

Given $w^\mu$ causal, let us consider the question of stability of the condensate background and write $w_\mu = (w_0,\vec w)$. First, suppose that $w$ is timelike, so $w^2 <0$. We can go to the condensate rest frame, so $\vec w = 0$. Then we have
\be
\mathcal{L}_\varphi = \left[-P'(w^2)+2P''(w^2)w_0^2\right] \dot \varphi^2  + P'(w^2) (\partial_i \varphi)^2.
\ee
If $P'(w^2) > 0$, there are ghosts in theory, resulting in a quantum mechanical pair-production instability \cite{Rubakov}. We thus conclude that $P'(w^2) \leq 0$ if $w^2 < 0$.  If $w$ is null, then we simply have $P'(0) = -1/2$. Hence, stability guarantees that $P'(w^2)$ is always nonpositive. Since in the $w^2=0$ case ${\cal L}_\varphi$ is trivial, we hereafter take $w$ to be timelike.

Let us now apply the causality bound \eqref{eq:causalbound} to the $n$th-order $P(X)$ theory, where all the $\lambda_i$ are negligible at leading order for $1<i<n$. By taking $w^2$ sufficiently small, we guarantee that $P''(w^2)$ is dominated by the $X^n$ term, which we take to be nonzero. We have
\be
P''(w^2) = n(n-1)(w^2)^{n-2} \lambda_n, 
\ee
so since $w^2 < 0$, 
\be
\begin{aligned}
\lambda_n > 0 & \text{ if $n$ is even},\\
\lambda_n < 0 & \text{ if $n$ is odd}.
\end{aligned}
\ee

\section{Bounds from Unitarity}\label{sec:Unitarity}

Let us again consider the $n$th-order $P(X)$ theory in which the first higher-dimension operator with nontrivial coefficient is $X^n$. For such a theory, we can consider a family of tree-level completions of the $X^s$ operator that takes the form of some combination of operators ${\cal O}_j$, where
\be 
\mathcal{O}_{j}=g_{j}\chi_{\mu_{1}\cdots\mu_{j}}\partial^{\mu_{1}}\phi\cdots\partial^{\mu_{j}}\phi. \label{eq:Os}
\ee
We generate $X^{s}$ whenever there is some part of $\chi_{\mu_{1}\cdots\mu_{j}}$ and $\chi_{\mu_{1}\cdots\mu_{k}}$ that are the same field (up to some extraneous metrics) for $j+k=s$. The coupling of $X^{s}$ will thus receive contributions that go as $g_{j}g_{k}$ for $j+k=s$. Of course, in that case the $X^{j}$ operator is also generated via the exchange of a $\chi_{\mu_{1}\cdots\mu_{j}}$ between two of the $\mathcal{O}_j$ operators and similarly for $X^k$. Thus, in a theory in which the tree-level coefficients $\lambda_i$ for $X^i$ are negligible, in units of the cutoff, compared to $\lambda_n$ for $1<i< n$, we must consider a completion in which the $g_{i}$ coefficients vanish for $1<i< n$. In such an $n$th-order $P(X)$ theory, the $X^{n}$ operator is generated by integrating out $\chi_{\mu_{1}\cdots\mu_{n}}$,  joining two copies of $\mathcal{O}_{n}$.\footnote{We will not consider theories in which the $n$-point operators in the completion vanish on-shell, e.g., for a traceless, spin-two massive state $\chi^{(2)}_{\mu\nu}$, a coupling of the form $\chi^{(2)}_{\mu\nu}\eta^{\mu\nu}(\partial \phi)^2$. Completions comprised purely of such operators do not have poles in their forward amplitudes associated with the massive state going on-shell.\label{caveat}}

Let us consider the structure of our massive states $\chi_{\mu_{1}\cdots\mu_{n}}$. Without loss of generality, we can take $\chi$ to be symmetric on its indices, since the interaction with $\partial^{\mu_{1}}\phi\cdots\partial^{\mu_{n}}\phi$ effectively projects out any nonsymmetric component. We can split $\chi$ up into its traces and traceless components by defining
\be
\begin{aligned}
\chi_{\mu_{1}\cdots\mu_{n}}&=\chi_{\mu_{1}\cdots\mu_{n}}^{(n)}+\eta^{\phantom{()}}_{(\mu_{1}\mu_{2}}\chi_{\mu_{3}\cdots\mu_{n})}^{(n-2)}+\eta^{\phantom{()}}_{(\mu_{1}\mu_{2}}\eta^{\phantom{()}}_{\mu_{3}\mu_{4}}\chi_{\mu_{5}\cdots\mu_{n})}^{(n-4)}+\cdots \\&=\sum_{s=0}^{2\lfloor n/2\rfloor}\eta^{\phantom{()}}_{(\mu_{1}\mu_{2}}\cdots\eta^{\phantom{()}}_{\mu_{2s-1}\mu_{2s}}\chi_{\mu_{2s+1}\cdots\mu_{n})}^{(n-2s)}, \label{eq:chiform}
\end{aligned}
\ee
where parentheses around subscripts denotes normalized symmetrization, i.e., $n! \, T_{(\mu_{1}\cdots\mu_{n})}= (T_{\mu_{1}\cdots\mu_{n}} + \text{permutations})$. 

We will bound $\lambda_n$ via an argument involving the K\"all\'en-Lehmann form of the exact propagator for the $\chi$ states.

\subsection{All Massive Bosonic Higher-Spin Propagators in Arbitrary $D$}

We now build the propagator numerator for $\chi_{\mu_{1}\cdots\mu_{s}}^{(s)}$. This is a canonical higher-spin state, that is, a symmetric tensorial rank-$s$ representation of the $SO(D-1)$ little group for a massive state in $D$ dimensions.\footnote{We will derive the unitary-gauge propagator numerator in the form of a Lorentz-covariant tensor; for a spin representation in $D=4$, see \Ref{Weinberg:1964cn}.} We require that $\chi_{\mu_{1}\cdots\mu_{s}}^{(s)}$ satisfy the Fierz-Pauli conditions \cite{Cortese:2013lda}, so that at leading order in $\chi^{(s)}$ in the equations of motion we have
\be
\begin{aligned}
\partial^{\mu_1} \chi_{\mu_{1}\cdots\mu_{s}}^{(s)} &= 0\\
\eta^{\mu_1 \mu_2} \chi_{\mu_{1}\cdots\mu_{s}}^{(s)}&=0.
\end{aligned} 
\ee 
Equivalently, the propagator numerator must be transverse and traceless on shell, when $k^2 = -m^2$, where $m$ is the mass of $\chi^{(s)}$.
We will write the propagator numerator for $\chi^{(s)}$ as $\Pi_{\mu_{1}\cdots\mu_{s}\nu_{1}\cdots\nu_{s}}$. Considering $k^{\mu_1} \Pi_{\mu_{1}\cdots\mu_{s}\nu_{1}\cdots\nu_{s}}$, the $\mu_1$ index with which $k$ contracts can either be on a metric, which by index symmetry on the $\mu$ indices we can write as $\eta_{\mu_1\mu_2}$, or another momentum $k_{\mu_1}$. We can therefore write $k^{\mu_1} \Pi_{\mu_{1}\cdots\mu_{s}\nu_{1}\cdots\nu_{s}} = Ak_{\mu_2} \psi_{\mu_3 \cdots \mu_s \nu_1 \cdots\nu_s} + B k^2 \omega_{\mu_2 \cdots \mu_s \nu_1 \cdots \nu_s}$, up to symmetrization, for some tensors $\psi$ and $\omega$ that are themselves built out of metrics and momenta. In order for this object to vanish on shell while leaving a nontrivial propagator, we must have $\omega_{\mu_2 \cdots \mu_s \nu_1 \cdots \nu_s}= k_{\mu_2} \psi_{\mu_3 \cdots \mu_s \nu_1 \cdots\nu_s}$ and $A = m^2 B$. That is, on-shell transversality requires that the propagator numerator be built out of the projector \cite{Weinberg:1969di}
\be 
\Pi_{\mu\nu}=\eta_{\mu\nu}+\frac{k_{\mu}k_{\nu}}{m^{2}}.
\ee

Without loss of generality, we can use symmetry of the propagator on the $\mu$ and $\nu$ indices separately, along with symmetry on the interchange of the sets of $\mu$ and $\nu$ indices, to write the general form the propagator numerator must take:
\be
\begin{aligned}\Pi_{\mu_{1}\cdots\mu_{s}}^{\qquad\nu_{1}\cdots\nu_{s}} & =\alpha_{0}^{(s)}\Pi_{(\mu_{1}}^{(\nu_{1}}\cdots\Pi_{\mu_{s})}^{\nu_{s})}+\alpha_{1}^{(s)}\Pi^{\phantom{()}}_{(\mu_{1}\mu_{2}}\Pi_{\phantom{()}}^{(\nu_{1}\nu_{2}}\Pi_{\mu_{3}}^{\nu_{3}}\cdots\Pi_{\mu_{s})}^{\nu_{s})}+\cdots\\
 & =\sum_{j=0}^{\lfloor s/2\rfloor}\alpha_{j}^{(s)}\Pi^{\phantom{()}}_{(\mu_{1}\mu_{2}}\cdots\Pi^{\phantom{()}}_{\mu_{2j-1}\mu_{2j}}\Pi_{\phantom{()}}^{(\nu_{1}\nu_{2}}\cdots\Pi_{\phantom{()}}^{\nu_{2j-1}\nu_{2j}}\Pi_{\mu_{2j+1}}^{\nu_{2j+1}}\cdots\Pi_{\mu_{s})}^{\nu_{s})}.\label{eq:propform}
\end{aligned} 
\ee
That is, if $s$ is even, the final term is $\Pi_{(\mu_{1}\mu_{2}}\cdots\Pi_{\mu_{s-1}\mu_{s})}\Pi^{(\nu_{1}\nu_{2}}\cdots\Pi^{\nu_{s-1}\nu_{s})}$, while if $s$ is odd, the final term is $\Pi_{(\mu_{1}\mu_{2}}\cdots\Pi_{\mu_{s-2}\mu_{s-1}}\Pi^{(\nu_{1}\nu_{2}}\cdots\Pi^{\nu_{s-2}\nu_{s-1}}\Pi_{\mu_{s})}^{\nu_{s})}$. Next, we enforce the tracelessness condition, which requires that $\eta^{\mu_{s-1}\mu_{s}}\Pi_{\mu_{1}\cdots\mu_{s}\nu_{1}\cdots\nu_{s}}=0$ on shell. We note that, on shell, $\eta^{\mu\nu}\Pi_{\mu\nu}=D-1$ and $\Pi_{\mu\alpha}\eta^{\alpha\beta}\Pi_{\beta\nu}=\Pi_{\mu\nu}$. We find
\be
\begin{aligned}&\eta^{\mu_{s-1}\mu_{s}}\Pi_{\mu_{1}\cdots\mu_{s}}^{\qquad\nu_{1}\cdots\nu_{s}} \stackrel{\text{on-shell}}{=}\\&\hspace{2cm} \frac{2}{s(s-1)}\sum_{j=0}^{\lfloor s/2\rfloor-1}\gamma_j^{(s)}\Pi^{\phantom{()}}_{(\mu_{1}\mu_{2}}\cdots\Pi^{\phantom{()}}_{\mu_{2j-1}\mu_{2j}}\Pi_{\phantom{()}}^{(\nu_{1}\nu_{2}}\cdots\Pi_{\phantom{()}}^{\nu_{2j+1}\nu_{2j+2}}\Pi_{\mu_{2j+1}}^{\nu_{2j+3}}\cdots\Pi_{\mu_{s-2})}^{\nu_{s})},
\end{aligned}
\ee
where, taking careful account of the combinatorics,
\be
\gamma_j^{(s)} =  \left(\begin{array}{c}
s-2j\\
2
\end{array}\right)\alpha_{j}^{(s)}+\left[(j+1)(D-1)+4\left(\begin{array}{c}
j+1\\
2
\end{array}\right)+2(j+1)(s-2j-2)\right]\alpha_{j+1}^{(s)}.
\ee
To enforce tracelessness, we thus require that each $\gamma_j^{(s)} = 0$, so
\be
\left(\begin{array}{c}
s-2j\\
2
\end{array}\right)\alpha_{j}^{(s)}+(j+1)(D-5+2s-2j)\alpha_{j+1}^{(s)}=0,\qquad 0\leq j\leq\lfloor s/2\rfloor-1. 
\ee
That is,
\be 
\begin{aligned}
\alpha_{j}^{(s)}&=(-1)^{j}\alpha_{0}^{(s)}\prod_{k=0}^{j-1}\frac{\left(\begin{array}{c} 
s-2k\\
2
\end{array}\right)}{(k+1)\left(D-5+2s-2k\right)} \\&= \left(-\frac{1}{2}\right)^j \alpha_0^{(s)} \frac{s!}{j!(s-2j)!}\frac{(2s-2j+D-5)!!}{(2s+D-5)!!}.
\end{aligned}\label{eq:alphajs}
\ee

Now, we need to determine $\alpha_{0}^{(s)}$, equivalently, the overall normalization of the propagator. Let us first count the number of degrees of freedom in $\chi_{\mu_{1}\cdots\mu_{s}}^{(s)}$. A tensor that is symmetric on $s$ indices in $d$ dimensions will have 
\be
N_{s,d} = \frac{d(d+1)\cdots(d+s-1)}{s!}=\left(\begin{array}{c}
d+s-1\\
s
\end{array}\right) 
\ee
independent components. The transverse condition restricts us to setting $d=D-1$ (i.e., going to the rest frame, we must have only spatial components). Furthermore, the tracelessness condition removes $N_{s-2,d}$ components, so the number of independent components is
\be 
\begin{aligned}
N&=\left(\begin{array}{c}
D+s-2\\
s
\end{array}\right)-\left(\begin{array}{c}
D+s-4\\
s-2
\end{array}\right)\\&=\left(\begin{array}{c}
D-4+s\\
s
\end{array}\right)\left(1+\frac{2s}{D-3}\right),
\end{aligned}
\ee
which matches the counting of \Ref{Cortese:2013lda}. In $D=4$, this expression reduces to the expected $N=2s+1$. 

Unitarity implies that, on shell, the propagator numerator can be written as a sum over a tensor product of the physical polarization states,
\be 
\Pi_{\mu_1 \cdots \mu_s \nu_1 \cdots \nu_s} = \sum_a \varepsilon(a)_{\mu_1 \cdots \mu_s}\varepsilon(a)^*_{\nu_1 \cdots \nu_s},
\ee
where $\varepsilon(a)_{\mu_1 \cdots \mu_s}$ are the unit-normalized spin-$s$ polarization states and $a$ is a label for the different states, with $\varepsilon(a)_{\mu_1 \cdots \mu_s} \varepsilon(b)^{\mu_1 \cdots \mu_s} = \delta_{ab}$ \cite{Schwartz:2013pla}. Hence, the full trace $\Pi_{\mu_1\cdots \mu_s}^{\qquad \mu_1 \cdots \mu_s}$ of the massive propagator numerator counts the number of physical degrees of freedom. As one can verify by computation, for the propagator numerator given in \Eq{eq:propform}, with the $\alpha_j^{(s)}$ coefficients given in \Eq{eq:alphajs}, we have 
\be
\Pi_{\mu_1\cdots \mu_s}^{\qquad \mu_1 \cdots \mu_s} = N\alpha_0^{(s)}. 
\ee
Thus, $\alpha_0^{(s)} = 1$ in arbitrary dimension, for arbitrary integer spin. That is, the propagator numerator for all massive higher-spin bosons in arbitrary dimension is
\be
 \begin{aligned}\Pi_{\mu_{1}\cdots\mu_{s}}^{\qquad\nu_{1}\cdots\nu_{s}} &=\sum_{j=0}^{\lfloor s/2\rfloor}\left(-\frac{1}{2}\right)^j \frac{s!}{j!(s-2j)!}\frac{(2s-2j+D-5)!!}{(2s+D-5)!!}\times\\
&\hspace{2cm}\times \Pi^{\phantom{()}}_{(\mu_{1}\mu_{2}}\cdots\Pi^{\phantom{()}}_{\mu_{2j-1}\mu_{2j}}\Pi_{\phantom{()}}^{(\nu_{1}\nu_{2}}\cdots\Pi_{\phantom{()}}^{\nu_{2j-1}\nu_{2j}}\Pi_{\mu_{2j+1}\phantom{()}}^{\nu_{2j+1}\phantom{()}}\!\!\!\cdots\Pi_{\mu_{s})}^{\nu_{s})}.
\end{aligned} \label{eq:allprops}
\ee
In the special case of $D=4$, \Eq{eq:allprops} matches the result of \Ref{Miyamoto}.\footnote{See also Refs.~\cite{Hayashi,Singh} for $D=4$. While, Eq. (A.3) of \Ref{Nayak} gives an expression for the arbitrary-$D$ propagator, their coefficient contains an error inherited from Eq. (32) of \Ref{Singh}.}

For example, the propagator numerator for a massive vector is just $\Pi_{\mu\nu}$, while the propagator numerators for massive states of spin $2$, $3$, $4$, and $5$ are:
\be
\begin{aligned}
&\text{spin 2:}\qquad &\Pi_{\mu_1\mu_2}^{\;\;\;\;\;\;\;\;\nu_1\nu_2} &= \Pi_{(\mu_1}^{(\nu_1}\Pi_{\mu_2)}^{\nu_2)} - \frac{1}{D-1}\Pi_{\mu_1\mu_2}\Pi^{\nu_1 \nu_2}\\
&\text{spin 3:}\qquad &\Pi_{\mu_1\mu_2\mu_3}^{\;\;\;\;\;\;\;\;\;\;\;\;\nu_1\nu_2\nu_3}&=\Pi_{(\mu_1}^{(\nu_1}\Pi_{\mu_2\phantom{|}\!}^{\nu_2\phantom{|}\!}\Pi_{\mu_3)}^{\nu_3)}-\frac{3}{D+1}\Pi^{\phantom{(}}_{(\mu_1\mu_2}\Pi_{\phantom{(}}^{(\nu_1 \nu_2}\Pi_{\mu_3)}^{\nu_3)}\\
&\text{spin 4:}\qquad &\Pi_{\mu_1\mu_2\mu_3\mu_4}^{\;\;\;\;\;\;\;\;\;\;\;\;\;\;\;\;\nu_1\nu_2\nu_3\nu_4}&=\Pi_{(\mu_1}^{(\nu_1}\Pi_{\mu_2\phantom{(}}^{\nu_2\phantom{(}}\Pi_{\mu_3\phantom{|}\!}^{\nu_3\phantom{|}\!}\Pi_{\mu_4)}^{\nu_4)}-\frac{6}{D+3}\Pi_{(\mu_1\mu_2}^{\phantom{|}\!}\Pi_{\phantom{|}\!}^{(\nu_1 \nu_2}\Pi_{\mu_3\phantom{|}\!}^{\nu_3\phantom{|}\!}\Pi_{\mu_4)}^{\nu_4)}\\&&&\qquad+\frac{3}{(D+1)(D+3)}\Pi_{(\mu_1\mu_2}\Pi_{\mu_3\mu_4)}\Pi^{(\nu_1\nu_2}\Pi^{\nu_3\nu_4)} \\
&\text{spin 5:}\qquad&\Pi_{\mu_1\mu_2\mu_3\mu_4\mu_5}^{\;\;\;\;\;\;\;\;\;\;\;\;\;\;\;\;\;\;\;\;\nu_1\nu_2\nu_3\nu_4\nu_5}&= \Pi_{(\mu_1}^{(\nu_1}\Pi_{\mu_2\phantom{|}\!}^{\nu_2\phantom{|}\!}\Pi_{\mu_3\phantom{|}\!}^{\nu_3\phantom{|}\!}\Pi_{\mu_4\phantom{|}\!}^{\nu_4\phantom{|}\!}\Pi_{\mu_5)}^{\nu_5)} - \frac{10}{D+5} \Pi_{(\mu_1\mu_2}^{\phantom{|}\!}\Pi_{\phantom{|}\!}^{(\nu_1 \nu_2}\Pi_{\mu_3\phantom{|}\!}^{\nu_3\phantom{|}\!}\Pi_{\mu_4\phantom{|}\!}^{\nu_4\phantom{|}\!}\Pi_{\mu_5)}^{\nu_5)}\\&&&\qquad+\frac{15}{(D+3)(D+5)}\Pi^{\phantom{|}\!}_{(\mu_1\mu_2}\Pi^{\phantom{|}\!}_{\mu_3\mu_4}\Pi_{\phantom{|}\!}^{(\nu_1\nu_2}\Pi_{\phantom{|}\!}^{\nu_3\nu_4}\Pi_{\mu_5)}^{\nu_5)}.
\end{aligned}
\ee
For the spin-2 case, we see that we have recovered the usual form of the massive graviton propagator numerator in $D$ dimensions \cite{Hinterbichler:2011tt}.

\subsection{Bounds for $P(X)$}

We are now equipped to compute the contribution to the effective operator $X^{n}$ coming from integrating out $\chi_{\mu_{1}\cdots\mu_{n}}$ in a theory containing the operator $\mathcal{O}_{n}=g_{n}\chi_{\mu_{1}\cdots\mu_{n}}\partial^{\mu_{1}}\phi\cdots\partial^{\mu_{n}}\phi$. As accounted for in \Eq{eq:chiform}, $\chi_{\mu_1 \cdots \mu_n}$ may contain states of spin $n$, $n-2$, etc. all the way down to spin 0 or 1, accompanied by metric tensors to make up the other indices. The K\"all\'en-Lehmann form of the exact propagator for the spin-$s$ state $\chi_{\mu_{1}\cdots\mu_{s}}^{(s)}$ is 
\be
\langle\chi_{\mu_{1}\cdots\mu_{s}}^{(s)}(k)\chi_{\nu_{1}\cdots\nu_{s}}^{(s)}(k')\rangle=i(-1)^s\delta^{D}(k+k')\int_{0}^{\infty}\mathrm{d}\mu^{2}\frac{\rho^{(s)}(\mu^{2})}{-k^{2}-\mu^{2}+i\epsilon}\Pi_{\mu_{1}\cdots\mu_{s}\nu_{1}\cdots\nu_{s}}. \label{eq:exactprop}
\ee
The $\rho^{(s)}(\mu^{2})$ are the spectral densities, which are nonnegative by unitarity in a theory free of ghosts, since $\rho^{(s)}(\mu^2)$ can be written as a sum over the norms of the set of intermediate states. The $(-1)^s$ factor is present due to our choices of sign conventions and metric signature. Since we have been explicitly considering the generation of the $X^n$ operators at tree level via the exchange of the massive $\chi$ states, the spectral densities are simply convenient notation for a sum over delta functions, as in \Ref{Cheung:2016wjt}: $\rho^{(s)}(\mu^2) = \sum_i n_i \delta(\mu^2 -m_i^{(s)2})$, where $m_i^{(s)}$ are the masses of the spin-$s$ states, with degeneracy $n_i$.

Let us now formally integrate out $\chi_{\mu_1\cdots\mu_n}$, treating the full multiplet in \Eq{eq:chiform}. If we attach two of the ${\cal O}_n$ vertices from \Eq{eq:Os} to the exact propagator in \Eq{eq:exactprop} and then compute the effective operator at low energies by sending $k$ to zero, we can calculate the coefficient $\lambda_n$ of $X^n$:
\be
\lambda_n =  (-1)^n \frac{g_n^2}{2} \int_0^\infty \frac{\mathrm{d}\mu^2}{\mu^2} \sum_{s=0}^{2\lfloor n/2\rfloor} \sum_{j=0}^{\lfloor (n-2s)/2\rfloor} \alpha_j^{(n-2s)} \rho^{(n-2s)}(\mu^2).
\ee
Computing the sum, one finds
\be
\beta_s = \sum_{j=0}^{\lfloor s/2\rfloor} \alpha_j^{(s)} = \frac{(D-4+2\lfloor s/2\rfloor)!!(D-5+2\lceil s/2\rceil)!!}{(D-4)!!(D-5+2s)!!},
\ee
Thus, $\beta_s>0$ in $D\geq 2$ for all $s$. In terms of $\beta_s$, we have
\be 
\lambda_n = (-1)^n \frac{g_n^2}{2} \int_0^\infty \frac{\mathrm{d}\mu^2}{\mu^2} \sum_{s=0}^{2\lfloor n/2 \rfloor} \beta_{n-2s} \rho^{(n-2s)}(\mu^2).\label{eq:lambdabeta}
\ee
By hypothesis, $\lambda_n \neq 0$. As a result, nonnegativity of the spectral density means that \Eq{eq:lambdabeta} implies
\be
\begin{aligned}
\lambda_n > 0 & \text{ if $n$ is even},\\
\lambda_n < 0 & \text{ if $n$ is odd}
\end{aligned} 
\ee
for a tree-level ultraviolet completion of the form defined in \Eq{eq:Os}. It would be interesting to apply even more general versions of the spectral representation argument to accommodate the other types of tree-level completions mentioned in footnote~\ref{caveat}, as well as loop-level completions; for the purposes of the present work, we can view the results of this section as an exploration of how the bound in \Eq{eq:bd}, which we derive in \Secs{sec:Analyticity}{sec:Causality} from analyticity and causality, comes about in particular explicit ultraviolet completions.

\section{Challenges of More General Bounds}\label{sec:General}

Thus far we have focused primarily on $n$th-order $P(X)$ theories. In this section, we discuss the difficulties inherent to using analyticity of scattering amplitudes to bound more general $P(X)$ theories. For example, let us consider the calculation of the six-point amplitude for three-to-three scattering in the forward limit for the general $P(X)$ theory
\begin{align}
\mathcal{L} = -\frac{1}{2}(\partial \phi)^2 + \lambda_4(\partial \phi)^4 +\lambda_6(\partial \phi)^6 +\cdots.
\end{align}
The three-to-three amplitude is computed from Feynman diagrams of two topologies: a six-point contact diagram and a diagram with $\phi$ exchange between two four-point vertices:
\be
\begin{aligned}
{\cal A} =& \phantom{+}\frac{1}{8} \lambda_6\left( s_{12} s_{34} s_{56} +\text{permutations}\right) \\&-16\lambda_4^2 \left[\frac{(s_{12}s_{13} + s_{12}s_{23} + s_{13}s_{23})(s_{45}s_{46} + s_{45}s_{56} + s_{46}s_{56})}{s_{12} + s_{13}+s_{23}} + \text{other channels}\right],\label{eq:amp6pt}
\end{aligned}
\ee
where ``$+\;\text{permutations}$'' indicates the sum over the other $6!-1$ permutations of the labels $\{1,\ldots,6\}$, while ``$+\;\text{other channels}$'' indicates the sum over the other nine ways of dividing the labels into two groups of three.

If we choose forward kinematics, $p_1 = -p_4$, $p_2 = -p_5$, $p_3 = -p_6$, then many of the channels have on-shell exchanged momentum; for example, for the 124 channel, the exchanged momentum is $p_1 + p_2 + p_4 = p_2$. Thus, the amplitude in \Eq{eq:amp6pt} possesses singularities at strictly forward kinematics. These singularities persist even if we make the $\phi$ massive: in that case, the denominator of the propagator becomes $p^2 + m^2$, where $p$ is the exchanged momentum and $m$ is the $\phi$ mass, so when $p$ goes on-shell, the amplitude again is singular.

While it is possible to consider almost-forward kinematics and take the forward limit in such a way that the singularity in particular powers of $s$ (e.g., $s^2$) vanishes, it is not clear that such a procedure produces a reliable positivity bound. For example, the optical theorem is independent of the way in which the forward limit is taken, so the limit-dependence that would show up in the residue computed at small $s$ makes the dispersion relation ambiguous. This issue is similar to the subtleties involving the $t$-channel singularity in gravity amplitudes \cite{Adams:2006sv,Cheung:2014ega,Bellazzini:2015cra}. We leave the investigation of these issues and the search for analyticity bounds on more general $P(X)$ theories to future work.   

\section{The Legendre Transform}\label{sec:Legendre}

Since multiple infrared consistency tests point to the same bounds on effective field theory coefficients, it is worthwhile considering whether these bounds are related to other physics principles. In this section, we will show that the positivity bounds we have derived on the $P(X)$ theory are connected with the consistency of the formulation of the mechanics of the theory.

In particular, given a theory specified by a Lagrangian ${\cal L}[\partial_\mu \phi,\phi]$, the Hamiltonian of the theory is given by acting on ${\cal L}$ with the Legendre transform $*$:
\be 
{\cal H}[p_\mu,\phi] = ({\cal L}[\partial_\mu \phi,\phi])^* = \sup_{\partial_\mu \phi} p^\mu \partial_\mu \phi  - {\cal L}[\partial_\mu \phi,\phi].
\ee
The Legendre transform is well defined when ${\cal L}$ is a convex function of $\partial_\mu \phi$. In particular, in a consistent formulation of the mechanics of a system free of constraints,  acting with the Legendre transform twice brings us back to the Lagrangian, i.e., the Legendre transform is an involution:
\be
{\cal L}^{**} = {\cal L} .
\ee
Convexity of ${\cal L}$ with respect to $\partial_\mu \phi$ implies that the supremum in the Legendre transform occurs when 
\be
\frac{\delta}{\delta (\partial_\mu \phi)}(p^\mu \partial_\mu \phi - {\cal L}) = 0,
\ee
so $p^\mu$ ends up being fixed to its canonical value, $p^\mu = \delta {\cal L}/\delta (\partial_\mu \phi)$.

In order to apply the Legendre transform to the Hamiltonian, we treat $p_\mu$ as an independent variable and require
\be
{\cal H}^* = \sup_{p_\mu} p^\mu \partial_\mu \phi - {\cal H}[p_\mu,\phi].
\ee
Consistency of the definition of the Legendre transform, which requires ${\cal L}$ be convex, also implies convexity of ${\cal H}$, so the supremum again occurs at a local extremum and we have 
\be
\frac{\delta}{\delta p_\mu} (p^\mu \partial_\mu \phi - {\cal H}) = 0, 
\ee
that is,
\be
\partial_\mu \phi = \frac{\delta {\cal H}}{\delta p^\mu}. 
\ee
Substituting this solution back into the definition of ${\cal H}^*$ and assuming that we can write $\partial_\mu \phi$ as an explicit functional $\partial_\mu \phi[p_\mu]$ of the canonical momentum $p^\mu  = \delta {\cal L}/\delta(\partial_\mu \phi)$, we have
\be
\begin{aligned}
{\cal L}^{**} = {\cal  H}^* &= (p^\mu \partial_\mu \phi - {\cal H}[p_\mu,\phi])|_{\partial_\mu \phi = \frac{\delta {\cal H}}{\delta p^\mu}}
\\ & = p^\mu \partial_\mu \phi[p_\mu] - (p^\mu \partial_\mu \phi - {\cal L}[p_\mu,\phi]) \\& = {\cal L}[\partial_\mu \phi,\phi].
\end{aligned}
\ee
Thus, the involution property of the Legendre transform is guaranteed if $p^\mu[\partial_\mu \phi]$ is invertible as $\partial_\mu \phi[p^\mu]$. See \Ref{Aharonov:1969vu} for further discussion of the connections between this invertibility property and causality.

For the $P(X)$ theory, $p^\mu = 2 P'(X) \partial^\mu\phi$ as we have previously noted. Thus, the canonical momentum is a mapping from one Lorentzian vector space to another. That is, $p_\mu$ is invertible for $p_\mu$ in the image of $\partial_\mu \phi$ provided this mapping is injective, i.e., the mapping of $\partial_\mu \phi$ to its image under $p^\mu$ is a diffeomorphism. Note that the map identifies spacelike, timelike, or null $\partial_\mu \phi$ with spacelike, timelike, or null $p_\mu$, respectively, so these identifications can be considered separately and must each be a diffeomorphism.

For some subset $\Omega \subset\mathbb{R}^n$, a map $f:\,\Omega \rightarrow \mathbb{R}^n$ is a diffeomorphism from $\Omega$ to $f(\Omega)$ if $\mathrm{d}f$ is positive or negative definite on $\Omega$ \cite{injective}. That is, viewed as a matrix, the Jacobian
\be
J_{\mu\nu} = \frac{\delta^2 \mathcal{L}}{\delta(\partial^\mu \phi)\delta(\partial^\nu \phi)}=2\left(P'(X)\eta_{\mu\nu}+2P''(X)\partial_\mu \phi\partial_\nu \phi\right)
\ee
must be positive or negative definite. Note that for a condensate background, $J_{\mu\nu} = -\tilde{\eta}_{\mu\nu}$, the effective metric for fluctuations $\varphi$ defined in \Eq{eq:effectivemetric}. Now, we know that $\lim_{X\rightarrow 0} P'(X) = -1/2$, regardless of the sign of $X$, while $\lim_{X\rightarrow 0} P''(X) =0$. Hence, the involution property holds if $J_{\mu\nu}$ is negative definite for all nonzero $X$. This occurs if and only if all the eigenvalues of $J_{\mu\nu}$ are negative. In particular, the eigenvectors of $J_{\mu\nu}$ are $\partial_\mu \phi$, with eigenvalues
\be 
e(X)= 2P'(X) + 4XP''(X),\label{eq:e}
\ee
so the involution property holds if $e(X)$ is negative:
\be
P'(X) + 2XP''(X)<0.\label{eq:involutioncondition}
\ee
For a timelike condensate, this is equivalent to saying that the effective metric $\tilde{\eta}_{\mu\nu}$ has the correct signature (i.e., the same signature as $\eta_{\mu\nu}$). That is, if we consider the setup of a stable timelike condensate with $X<0$ and $P'(X)<0$, the causality bound in \Eq{eq:causalbound} implies that the condition in \Eq{eq:involutioncondition} holds, so the Legendre transform is an involution relating the Lagrangian and Hamiltonian.

As a final observation, we note that the involution property is related to the weak energy condition. Again taking a timelike condensate $w_\mu$ as in \Sec{sec:Causality}, the weak energy condition requires that $\overline{T}_{\mu\nu} w^\mu w^\nu \geq 0$. But from \Eq{eq:energymomentum}, $\overline{T}_{\mu\nu}w^\mu w^\nu =  P(w^2)w^2 -2P'(w^2)(w^2)^2$. Comparing with \Eq{eq:e}, we notice that
\be
\overline{T}_{\mu\nu}w^\mu w^\nu =  -\frac{w^2}{2}\int_0^{w^2} \mathrm{d}X\,e(X),
\ee
where the last inequality follows from \Eq{eq:involutioncondition}. Hence, the weak energy condition, which requires $T_{\mu\nu}t^\mu t^\nu \geq 0$ for all timelike $t^\mu$, implies $\int_0^{w^2} {\rm d}X\,e(X) > 0$ for $w^2 < 0$, which is the integral form of the requirement of involution of the Legendre transform.
Similarly, the causality bound $P''(X)\geq 0$ in \Eq{eq:causalbound} implies the dominant energy condition, which stipulates causality of the flux of energy-momentum seen by any inertial observer \cite{Adams:2006sv}.

\section{Conclusions}\label{sec:Conclusions}

In this paper, we have extended to higher-point terms the techniques of placing positivity bounds on higher-dimension operators in effective field theories using principles of infrared consistency. In the context of a theory polynomial in $X=(\partial\phi)^2$, we showed that in theories where the first nonnegligible higher-dimension operator is at $n$th order in $X$, these infrared consistency bounds imply that $\lambda_n > 0$ if $n$ is even and $\lambda_n < 0$ if $n$ is odd, in mostly-plus metric signature. 

We presented multiple different arguments for these bounds. In particular, we proved the bounds using analyticity of $2n$-point scattering amplitudes, as well as another proof using causality and the absence of superluminality in the low-energy theory. In a particular class of tree-level ultraviolet completions, we saw how these bounds arise from unitarity. By considering these lines of argument, we were able to extend useful techniques that will allow higher-point operators to be bounded in other theories. For example, we examined the additional kinematic freedom in the forward limit inherent to higher-point operators. We also exhibited a succinct derivation of the propagator numerators for all massive higher-spin bosons in arbitrary dimension, obtaining their form from symmetries and simple physical constraints alone.

Much work remains to be done to map out the space of possible low-energy effective field theories. In \Sec{sec:General}, we illustrated the challenges endemic to placing analyticity bounds on more general $P(X)$ theories due to kinematic singularities; these issues are similar in nature to the difficulties in addressing $t$-channel singularities in gravity theories discussed in Refs.~\cite{Adams:2006sv,Cheung:2014ega,Bellazzini:2015cra} and the challenge of proving the $a$-theorem in six dimensions discussed in \Ref{Elvang:2012st}. Further work on infrared consistency conditions for multipoint operators has the potential to further our understanding of these questions. 

Finally, elucidating the deep relationships among constraints on effective field theories is an important topic for future study. In this paper, we derived the same constraint from analyticity, unitarity, and causality and also showed how infrared constraints on the $P(X)$ action are related to the well-posedness of the Legendre transform relating the Lagrangian and Hamiltonian formulations of the theory. Infrared constraints such as these complement bounds obtainable from ultraviolet reasoning. A more complete understanding of the connections between ultraviolet and infrared within the swampland program remains a compelling topic for future work.
 
\begin{center} 
 {\bf Acknowledgments}
 \end{center}

 \noindent 

We thank Nima Arkani-Hamed, Brando Bellazzini, and Cliff Cheung for useful discussions and comments. V.C. and A.S.-M. are supported in part by the Berkeley Center for Theoretical Physics, by the National Science Foundation (award numbers 1214644, 1316783, and 1521446), by fqxi grant RFP3-1323, and by the US Department of Energy under Contract DE-AC02-05CH11231. G.N.R. is supported by the Miller Institute for Basic Research in Science at the University of California, Berkeley.

\bibliographystyle{utphys-modified}
\bibliography{P(X)}
\end{document}